\documentclass{PoS}
\usepackage{amsmath}
\usepackage{braket}
\usepackage{bm}
\usepackage{url}

\title{Charmonium-nucleon interactions from $2+1$ flavor lattice QCD}

\ShortTitle{Charmonium-nucleon interactions from $2+1$ flavor lattice QCD}

\author{
	\speaker{Takuya Sugiura}\\
        E-mail: \email{sugiura@rcnp.osaka-u.ac.jp}
}
\author{
	Yoichi Ikeda
}
\author{
	Noriyoshi Ishii \\
	Research Center for Nuclear Physics (RCNP), Osaka University
}

\abstract{
  The charmonium-nucleon interaction is studied by the time-dependent
  HAL QCD method. We use a larger lattice volume and the relativistic
  heavy quark action for charm quark to obtain less systematic errors
  than those in our previous study.  As a result, the sizable $J/\psi
  N$ hyperfine splitting is observed, indicating that the spin-spin
  interaction is important to understand this system
  quantitatively. No $J/\psi N$ or $\eta_c N$ bound state is observed
  below the thresholds as in the previous results.
}

\FullConference{The 36th Annual International Symposium on Lattice
  Field Theory - LATTICE2018\\ 22-28 July, 2018\\ Michigan State
  University, East Lansing, Michigan, USA.}

\begin{document}

\section{Introduction}

The LHCb collaboration has reported a peak structure in the $J/\psi p$
invariant mass spectrum of the weak decay process $\Lambda_b \to
J/\psi p K^-$~\cite{Aaij:2015tga}. The two resonances, $P_c(4380)$ and
$P_c(4450)$, are found in the mass spectrum with more than 9 standard
deviation significance by the fit with two Breit-Wigner resonances for
$P_c$'s and 14 known $\Lambda^*$ resonances.  The three preferred
spin-parity assignments are given as $(3/2^-,5/2^+)$, $(3/2^+,5/2^-)$,
$(5/2^+,3/2^-)$ for $P_c(4380)$ and $P_c(4450)$, respectively.
Furthermore, the LHCb recently has analyzed nine times more data than
those in the 2015 analysis and found a new resonance $P_c(4312)$ and
two narrow overlapping peaks $P_c(4440)$ and $P_c(4457)$ instead of
$P_c(4450)$~\cite{LHCb_press}.  Since the $P_c$ states are found in
the $J/\psi N$ mass spectrum above the threshold, they are naturally
expected to have $uudc\bar{c}$ flavor structure, {\it i.e.}, the
pentaquarks.  The discovery of the hidden-charm pentaquarks attracts
much attention to explain the origin of these resonances.  Since
experimental data for interactions between charmed hadrons is very
scarce, effective theories will suffer from large ambiguity.  In this
situation, lattice QCD first-principle calculation of these
interactions is of significant importance.  Lattice QCD also has an
advantage over experimental studies, since we can directly calculate
the $J/\psi N$ scattering, without requiring the spectator $K^-$.

A possible explanation of $P_c$ is that they are loosely bound
$\bar{D}^{*} \Sigma_c$ and $\bar{D}^{*} \Sigma_c^*$
states~\cite{Roca:2016tdh, Shimizu:2017xrg}. The interaction between
these states is supplied by meson exchange forces. Another picture is
proposed in Refs.~\cite{Eides:2015dtr,Eides:2017xnt}, where the
possibility of having $P_c(4450)$ as a deeply bound narrow
$\psi(2S)$-nucleon bound state is discussed. In this scenario,
$\psi(2S)$ is assumed to be a Coulomb bound state, so that the
description of the QCD van der Waals force is
applicable~\cite{Peskin}. The strength of the QCD van der Waals force
is determined by the chromoelectric polarizability of the charmonium,
which is approximately proportional to inverse cube of the charmonium
radius.  Thus the $\psi(2S) N$ interaction can be very strong and
possibly explain as large as $176\mathrm{MeV}$ binding energy of
$P_c(4450)$ as a narrow $\psi(2S) N$ bound state.  Although this
picture is theoretically interesting, we need to be careful to obtain
quantitative understanding what $P_c$'s are like.  In
Ref.~\cite{Polyakov:2018aey}, results of our previous
calculation~\cite{Sugiura:2017vks} have been used for a rough, but
quantitative evaluation of the $J/\psi$ chromoelectric polarizability.
In this present paper, we study the $J/\psi N$ and $\eta_c N$
scattering in lattice QCD with better statistics and systematics than
those in Ref.~\cite{Sugiura:2017vks}.

As before, we employ the method developed by the HAL QCD
collaboration~\cite{Ishii:2006ec,HALQCD:2012aa}. The HAL QCD method can be
straightforwardly extended to coupled-channel systems~\cite{Aoki:2012tk};
this is an important advantage over L\"uscher's finite volume
method~\cite{Luscher:1990ux}. We have improved the calculations by
Kawanai and Sasaki~\cite{Kawanai:2010ev} by using the time-dependent
HAL QCD method and found that the charmonium-nucleon interaction has
significantly stronger attraction than their results.  We have also
found significant hyperfine splitting of the $J/\psi N$ interaction
between the $J=1/2$ and $J=3/2$ states.  This paper is organized as
follows. In Sec.~\ref{sec:method}, we briefly introduce the HAL QCD
(both original and time-dependent), mentioning possible origins of
systematic errors and necessary conditions to be satisfied.  In
Sec.~\ref{sec:setup} we explain the lattice QCD
setup. Sec.~\ref{sec:results} shows our results for the $J/\psi N$ and
$\eta_c N$ interactions. We discuss importance of using the
time-dependent method by taking a close look at the terms appearing by
temporal derivatives.  We summarize our conclusions in
Sec.~\ref{sec:conclusions}.

\section{Method \label{sec:method}}

The charmonium-nucleon four point function is defined as
\begin{align}
  R(\bm{x}-\bm{y},t-t_0)
  &=
  \Braket{0|T \phi(\bm{x},t) N(\bm{y},t) \overline{\mathcal{J}}(t_0)|0}
  \times e^{(m_N + m_\phi)t}
  \\
  \label{eq:Rcorr_decomposition}
  &=
  \sum_n A_n \psi_n(\bm{r}) e^{-\Delta W_n t},
\end{align}
where $\phi(\bm{y},t)$ and $N(\bm{x},t)$ are local interpolating
operators for charmonium (either $J/\psi$ or $\eta_c$) and nucleon,
respectively, and $\mathcal{J}(t_0)$ is the corresponding wall-source
operator. The masses of $\phi$ and $N$ are denoted as $m_\phi$ and
$m_N$.  The second line can be derived by inserting the complete set
of the QCD eigenstates $\bm{1}=\sum_n \ket{n}\bra{n}$ and by defining
$\psi_n(\bm{r})=\Braket{0|\phi(\bm{r}+\bm{x},0)N(\bm{x},0)|n}$,
$A_n=\Braket{n|\overline{\mathcal{J}}(0)|0}$, the $n$-th eigenvalue
$W_n = \sqrt{k_n^2 + m_\phi^2} + \sqrt{k_n^2 + m_N^2}$ and $\Delta W_n
= W_n - W_0$.  $\psi_n(\bm{r})$ is called the equal-time
Nambu-Bethe-Salpeter~(NBS) wave function, as it satisfies the
Helmholtz equation in the long-distance limit $|\bm{r}|\to \infty$.
The energy-independent and non-local potential
$U(\bm{r},\bm{r}^\prime)$ is defined by~\cite{Ishii:2006ec}
\begin{align}
\label{eq:hal_org}
  \left(
    \frac{\bm{\nabla}^2}{2\mu} + E_n
  \right)
  \psi_n(\bm{r})
  =
  \int d^3 \bm{r}^\prime
  U(\bm{r},\bm{r}^\prime)
  \psi_n(\bm{r}^\prime),
\end{align}
where $\mu=1/(1/m_\phi+1/m_N)$ is the reduced mass and
$E_n=k_n^2/(2\mu)$.  In the original HAL QCD method, one solves
Eq.~\eqref{eq:hal_org} for the potential $U$; for that one has to
extract the NBS wave function for the ground state by assuming the
{\it ground-state saturation}, {\it i.e.}, $R(\bm{r},t) \simeq A_0
\psi_0(\bm{r}) e^{-\Delta W_0t}$ at large $t$. In actual situations,
ground-state saturation is hard to achieve because the signal-to-noise
ratio becomes exponentially bad for large $t$.  An alternative method
has been developed in Ref.~\cite{HALQCD:2012aa}. The same potential
$U(\bm{r},\bm{r}^\prime)$ satisfies the following time-dependent
Schr\"odinger-like equation upto $\mathcal{O}(k_n^6)$:
\begin{align}
\label{eq:hal_tdep}
  \left(
      \frac{\bm{\nabla}^2}{2\mu}
    - \frac{\partial}{\partial t}
    + \frac{1+3\delta^2}{8\mu} \frac{\partial^2}{\partial t^2}
  \right)
  R(\bm{r},t)  
  =
  \int d^3 \bm{r}^\prime
  U(\bm{r},\bm{r}^\prime)
  R(\bm{r}^\prime, t),
\end{align}
where $\delta=(m_\phi-m_N)(m_\phi+m_N)$. Eq.~\eqref{eq:hal_tdep} is
satisfied upto inelastic state contributions; in other words, one
requires that the four-point function is dominated by the states with
energy below the threshold energy of the next channel $W_{th}$
(elastic-state saturation).  In our lattice setup with heavy pion
mass, that is $W_{th}=m_{D}+m_{\Sigma_c}$.  The elastic-state
saturation is much easier to achieve than the ground-state saturation,
and one can reliably extract the potentials by
Eq.~\eqref{eq:hal_tdep}.
By taking the lowest-order term of the derivative expansion, {\it
  i.e.} $U(\bm{r},\bm{r}^\prime)\simeq V_{\text{eff}}(r) \delta^3
(\bm{r}-\bm{r}^\prime)$~\footnote{Note that the $J/\psi N$ potential
  consists of the central, the spin-spin, and two tensor forces at the
  lowest order of the derivative expansion. We can determine each of
  these four forces by using linearly-independent correlation
  functions for four different states. Then we find that the S-D wave
  orbital angular momentum mixing due to the tensor forces is much
  smaller than the central and the spin-spin forces, so that we can
  tentatively work without considering S-D mixing explicitly.  The
  detailed results of the spin-dependent forces are presented
  elsewhere.}, one can calculate the effective central potential as
\begin{align}
  V_{\text{eff}}(r)
  =
    \frac{1}{2\mu} \frac{\bm{\nabla}^2 R(\bm{r},t)}{R(\bm{r},t)}
  - \frac{\partial_t R(\bm{r},t)}{R(\bm{r},t)}
  + \frac{1+3\delta^2}{8\mu} \frac{\partial_t^2 R(\bm{r},t)}{R(\bm{r},t)}.
\end{align}
Although the right-hand side seems to depend on $t$, the left-hand
side does not as far as the elastic-state saturation is achieved and the
derivative expansion converges at this order. Therefore, by seeing the
$t$-independence of $V_{\text{eff}}(r)$ we can check the necessary
condition for our assumptions: the elastic-state saturation and the
convergence of the derivative expansion. Moreover, when the
ground-state saturation is achieved, the time-derivative terms do not
depend on the spatial distance $r$; the $(\partial_t R)/R$ term
roughly corresponds to $E_n$ and the $(\partial_t^2 R)/R$ term
corresponds to non-relativistic correction of
$\mathcal{O}(k_n^4)$. Although we do not need the ground-state
saturation anymore, this property will help us understand our results
in comparison to previous calculations based on the original HAL QCD
method.

In this paper, we neglect the coupling of $J/\psi N$ and $\eta_c N$,
since such a charmonium spin-flipping transition is suppressed due to
the large charm quark mass.  We also neglect the OZI-suppressed
$c\bar{c}$ annihilation diagrams to calculate the four-point function
in Eq.~\eqref{eq:Rcorr_decomposition}.

\section{Simulation Setup \label{sec:setup}}

We employ the 2+1 flavor QCD gauge configurations on a $32^3\times 64$
lattice, which is generated by the PACS-CS
collaboration~\cite{PACS-CS} with the renormalization group improved
gauge action at $\beta=1.9$ and the non-perturbatively
$\mathcal{O}(a)$ improved Wilson quark action at $c_{SW} = 1.715$.
The corresponding lattice spacing is $a=0.0907(13)\mathrm{fm}$ and the
spatial volume is $L=(2.90\mathrm{fm})^3$. We use the hopping
parameters $\kappa_{\text{ud}}=0.13700$ and
$\kappa_{\text{s}}=0.13640$. For the charm quark we employ the
Tsukuba-type relativistic heavy quark (RHQ) action to remove the
leading and next-to-leading order cutoff errors~\cite{Aoki:2001ra}.
We use the RHQ parameters determined in Ref.~\cite{Namekawa:2011wt}
such that physical hadron properties are reproduced at physical quark
masses.  The periodic boundary condition is imposed in the spatial
directions, while the Dirichlet boundary condition is imposed in the
temporal direction at $(t-t_0)/a=32$ to prevent inverse propagation.
We use 32 different source positions $t_0$ and take their average.

\section{Results and Discussion \label{sec:results}}

\begin{figure}[tbh]
  \begin{minipage}{0.5\hsize}
  \centering
  \includegraphics[width=\hsize,bb=0 0 792 612]{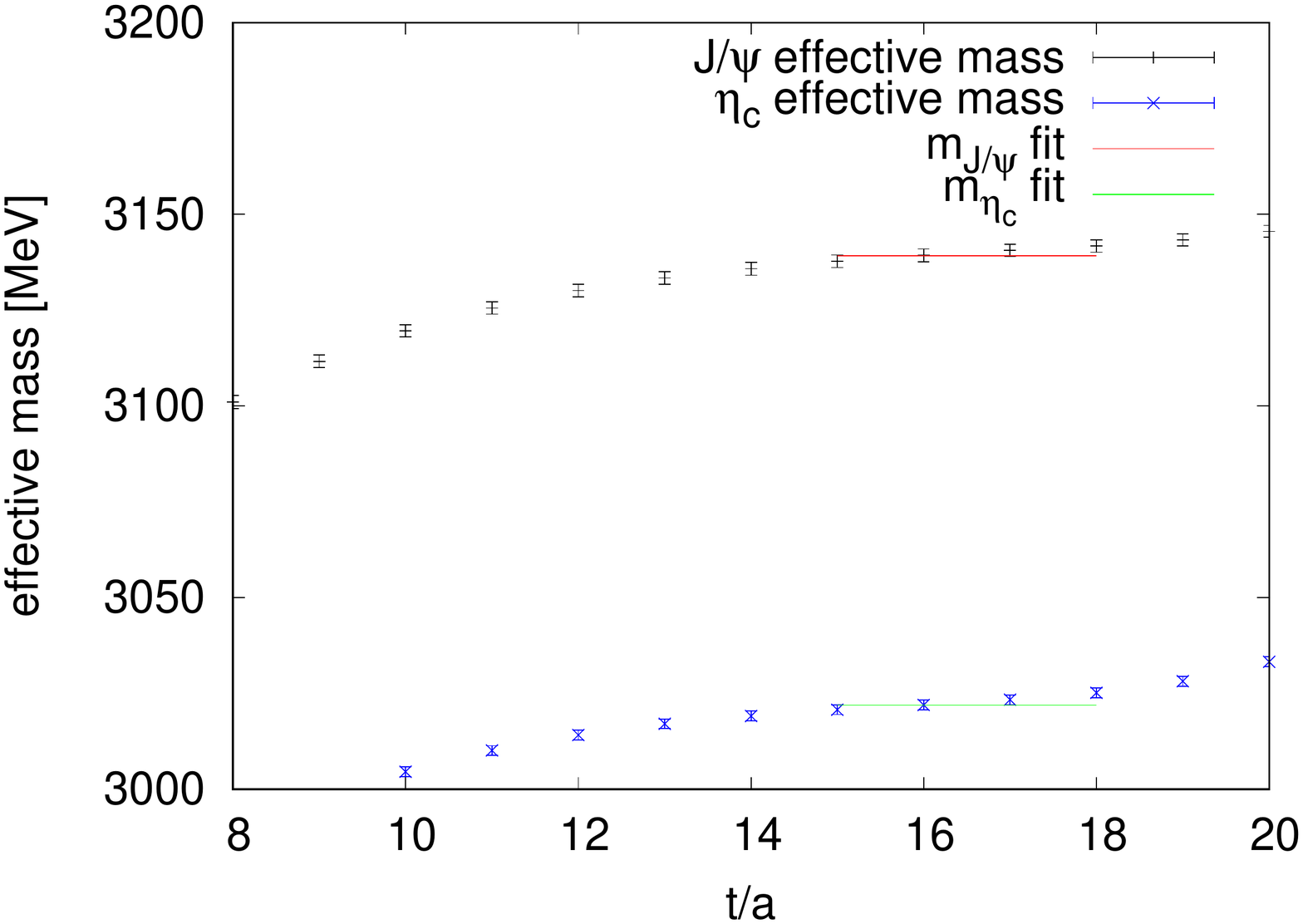}
  \end{minipage}
  \begin{minipage}{0.5\hsize}
  \centering
  \includegraphics[width=\hsize,bb=0 0 792 612]{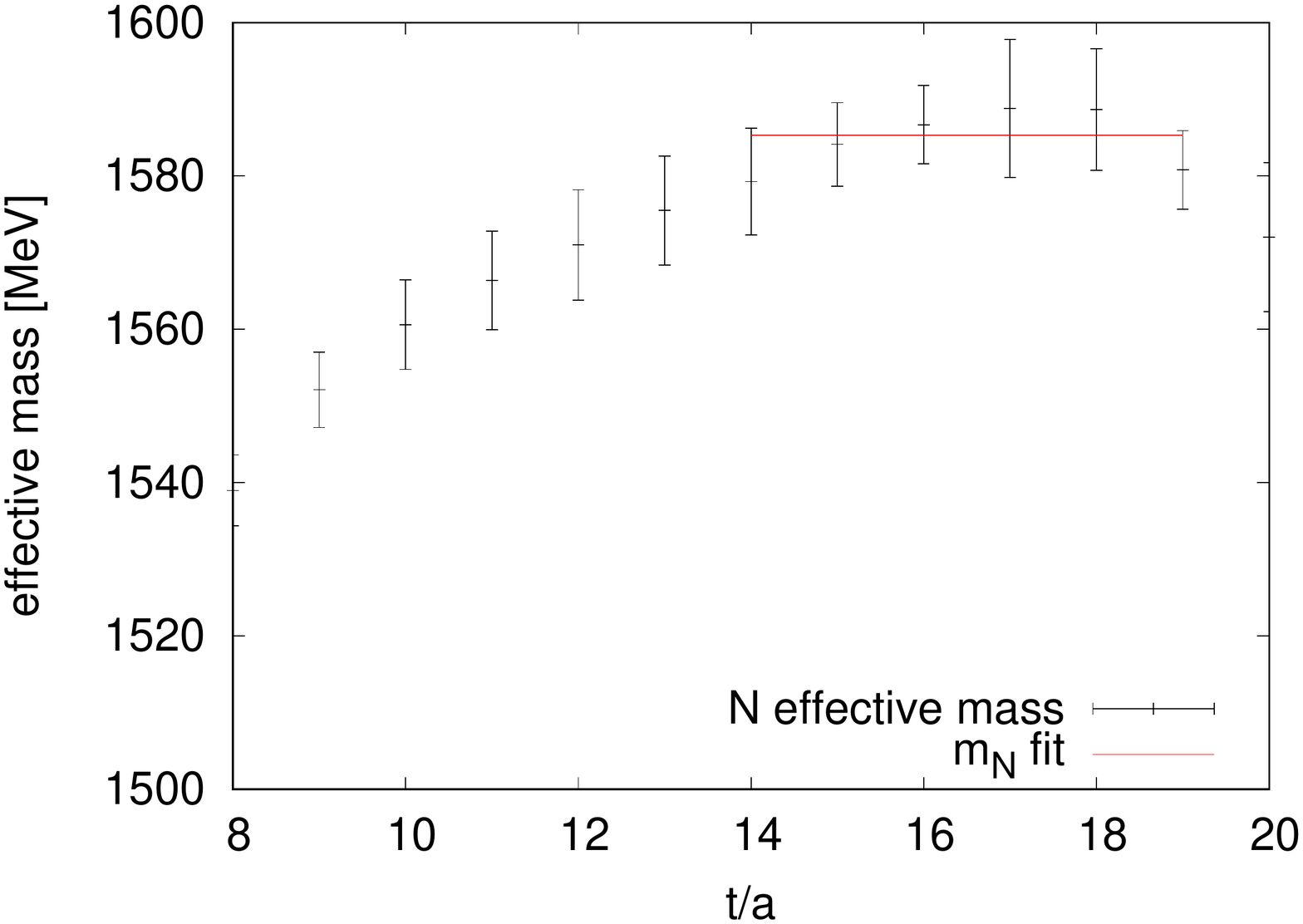}
  \end{minipage}
  \caption{\label{fig:effmass}
    Effective masses for (left) $J/\psi$ and $\eta_c$ and (right) nucleon.
  }
\end{figure}

Shown in Fig.~\ref{fig:effmass} are the effective masses
$m_{\text{eff}}(t)=(1/a) \ln\left(C_2(t)/C_2(t+a)\right)$ for $J/\psi$,
$\eta_c$, and $N$ calculated from the corresponding hadron two-point
functions $C_2(t)$.  The masses are evaluated by fitting the two-point
functions by a single exponential in the plateau region of the
effective mass plots.  We find that $m_{J/\psi} =
3139(12)\mathrm{MeV}$ ($(t-t_0)/a=15-18$),
$m_{\eta_c}=3022(9)\mathrm{MeV}$ ($(t-t_0)/a=15-18$), and $m_N
=1585(16)\mathrm{MeV}$ ($(t-t_0)/a=14-19$), where the numbers in the
first parentheses are statistical errors, and the second parentheses
indicate the fit ranges.  Similar analysis for pion shows $m_\pi =
700(2)\mathrm{MeV}$ ($(t-t_0)/a=11-18$).

\begin{figure}[t]
  \begin{minipage}{0.47\hsize}
    \centering
    \mbox{\raisebox{0mm}{\includegraphics[width=\hsize,bb=0 0 792 612]{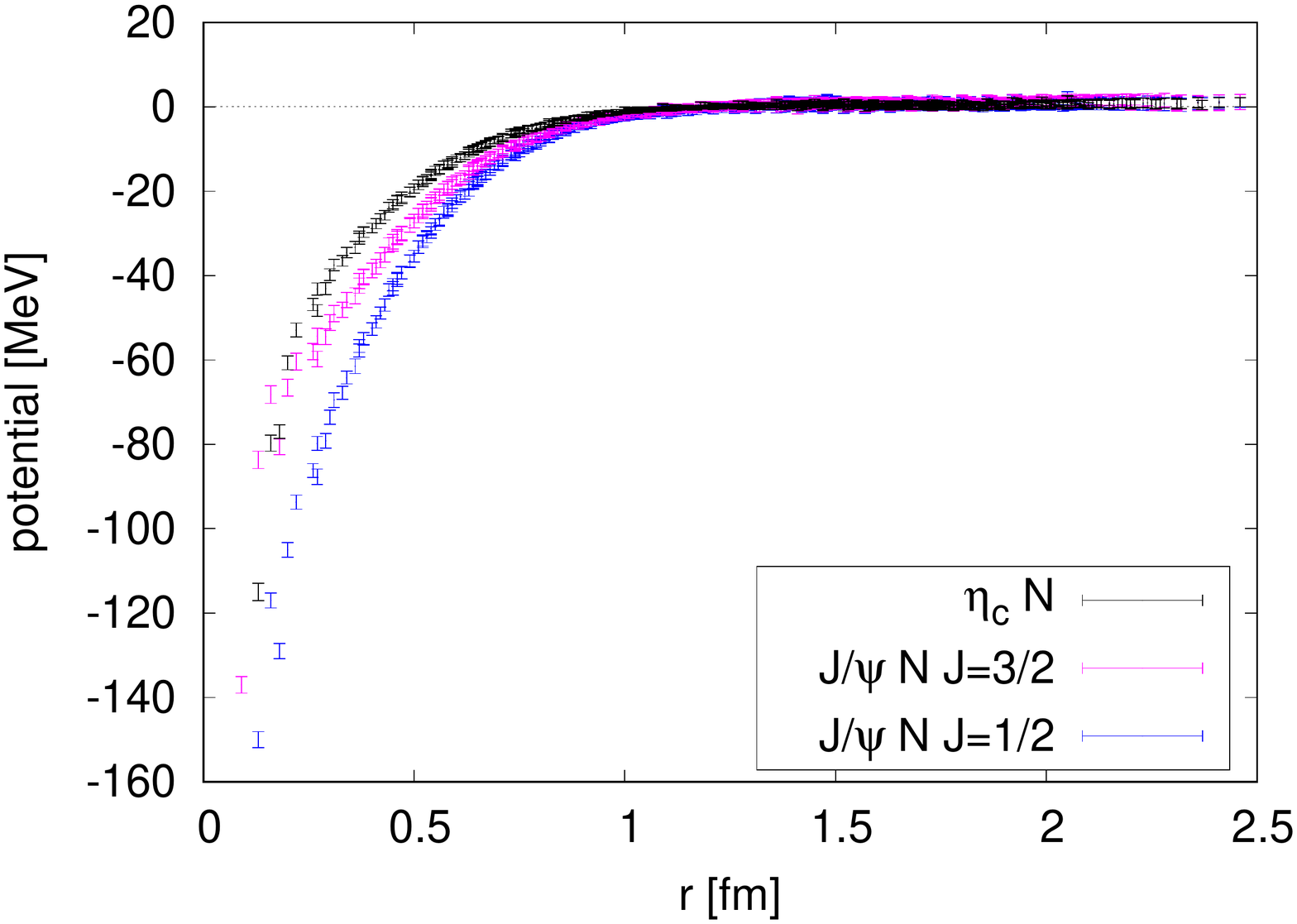}}}
    \caption{\label{fig:veff_all}
      The S-wave effective central potentials for (blue) $J/\psi N$
      ($J=1/2$), (magenta) $J/\psi N$ ($J=3/2$), and (black) $\eta_c N$
      evaluated at $(t-t_0)/a=15$.
    }
  \end{minipage}
  \hspace{0.05\hsize}
  \begin{minipage}{0.47\hsize}
    \centering
    \mbox{\raisebox{0mm}{\includegraphics[width=\hsize,bb=0 0 792 612]{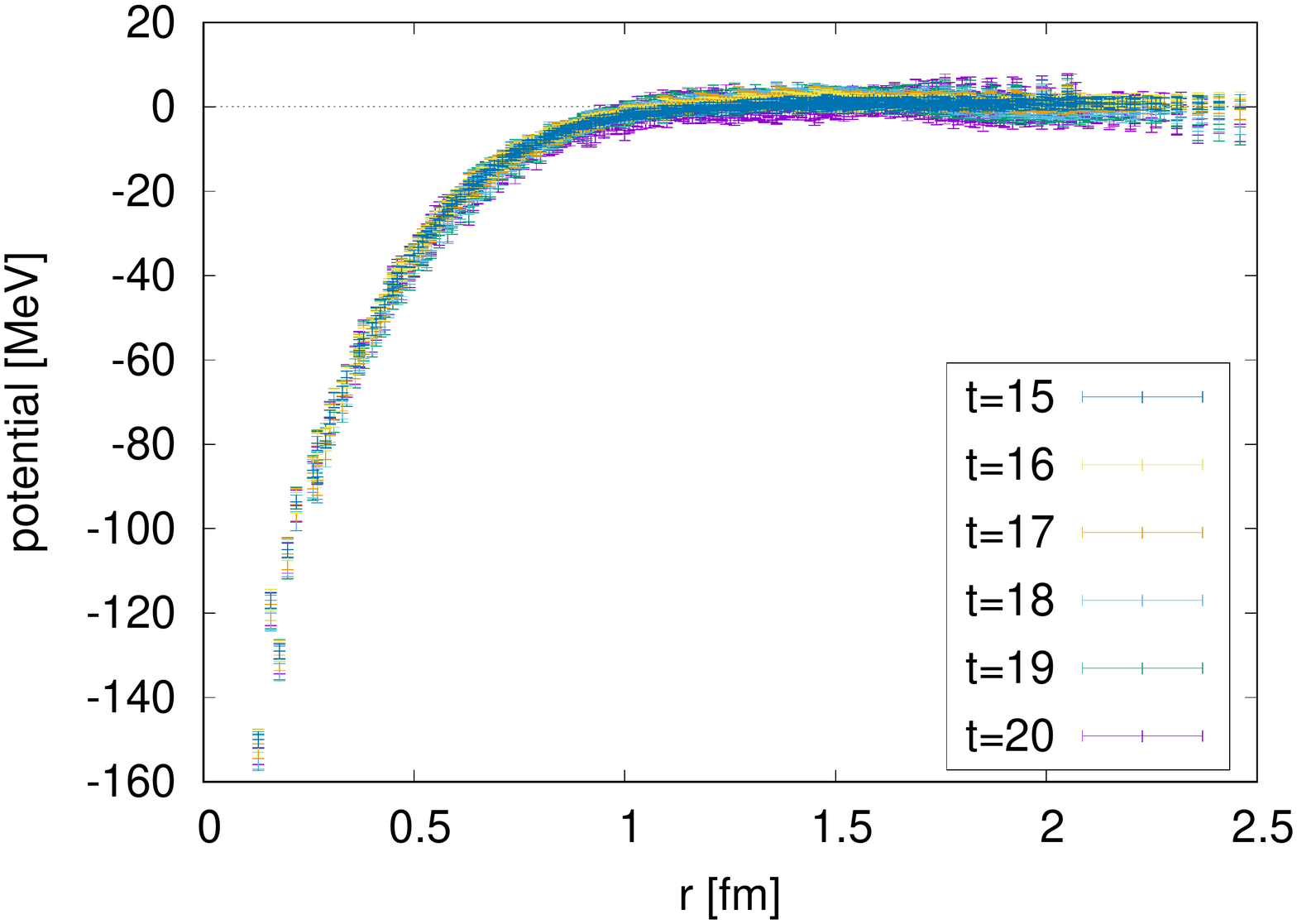}}}
    \caption{\label{fig:veff_tdep}
      The Euclidean time dependence of the $J/\psi N$ ($J=1/2$)
      effective central potential.
    }
  \end{minipage}
\end{figure}

In Fig.~\ref{fig:veff_all}, we show the S-wave effective central
potentials for $J/\psi N$ ($J=1/2$), $J/\psi N$ ($J=3/2$), and $\eta_c
N$ evaluated at $(t-t_0)/a=15$.  We see that all of them are
attractive overall and the $J/\psi N$ interactions (both $J=1/2$ and
$J=3/2$) are stronger than the $\eta_c N$ interaction, as already
reported previously~\cite{Sugiura:2017vks,Kawanai:2010ev}.
We also observe the hyperfine splitting of $J/\psi N$ interactions;
the $J=1/2$ state has stronger attraction than the $J=3/2$ state.  In
Ref.~\cite{Sugiura:2017vks}, the hyperfine splitting is much
smaller than the difference between $J/\psi N$ and $\eta_c N$
potentials, and than the absolute value of the $J/\psi N$ potential.
In this study we use a larger lattice with volume $La=2.90\mathrm{fm}$
than the one in Ref.~\cite{Sugiura:2017vks} with
$La=1.93\mathrm{fm}$.  Also, in this study we employ the RHQ action
for charm quark to reduce discretization errors, while in
Ref.~\cite{Sugiura:2017vks} we do not.  Since the $J/\psi N$
interaction is short-ranged such that $V_{\text{eff}}(r)\simeq 0$ for
$r>1\mathrm{fm}$, the finite volume effect should not be important.
Thus it seems that we can now see the $J/\psi N$ hyperfine splitting
clearly thanks to the use of the RHQ action.  The origin of the
hyperfine splitting is the spin-spin force, which arises due to
interference of the chromoelectric dipole and the chromomagnetic
quadrupole in the QCD van der Waals framework.

In order for the potential to be correct, we have to confirm that the
elastic-state saturation for the R-correlator in
Eq.~\eqref{eq:Rcorr_decomposition} and that the derivative expansion
converges at this order.  These assumptions can be checked by seeing
the $t$-independence of the resulting potential.
To do this~\footnote{ Rigorously speaking, the lowest-order $J/\psi N$
  potential consists of the central, spin-spin, and two tensor forces,
  so that the $t$-independence of each of these forces should be
  checked.  Here we show the $t$-independence of the effective central
  potential instead, but we have also confirmed the $t$-independences
  of the four forces explicitly.  },
in Fig.~\ref{fig:veff_tdep}, we show the $J/\psi N$ ($J=1/2$)
effective central potentials evaluated at different $t-t_0$.  We see
no significant $t$-dependence in the potential from $(t-t_0)/a=15$ to
$20$, so that our necessary condition is satisfied at $(t-t_0)/a=15$
(corresponding to Fig.~\ref{fig:veff_all}).  We can also check the
$t$-independence of the $J/\psi N$ ($J=3/2$) and $\eta_c N$ potentials.

\begin{figure}[htb]
  \begin{minipage}{0.5\hsize}
    \centering
    \includegraphics[width=\hsize,bb=0 0 792 612]{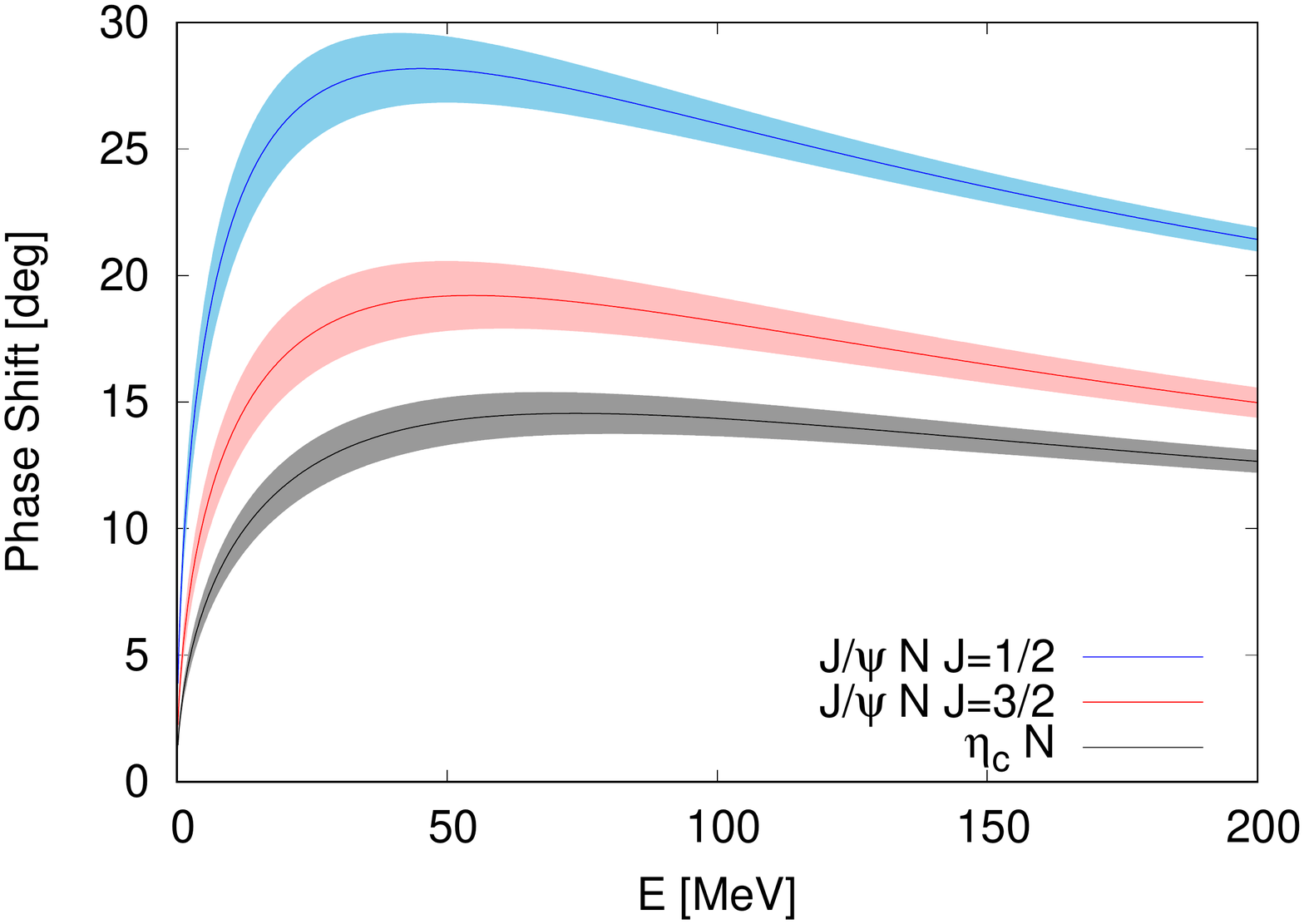}
    \caption{\label{fig:phase_shift}
      S-wave scattering phase shifts calculated from the potentials in
      Fig.~\ref{fig:veff_all} as a function of the center-of-mass energy.
    }
  \end{minipage}
  \begin{minipage}{0.5\hsize}
    \centering
    \makeatletter
    \def\@captype{table}
    \makeatother
    \caption{
      \label{table:scattering_length}
      The scattering length and the effective range from the effective central potentials.
    }
    \begin{tabular}{|c|c|c|} \hline
      channel               &   $a [\mathrm{fm}]$   &   $r [\mathrm{fm}]$  \\ \hline
      $J/\psi N$, $J=1/2$   &   $0.656 \pm 0.071$   &   $1.105 \pm 0.016$  \\ \hline
      $J/\psi N$, $J=3/2$   &   $0.380 \pm 0.048$   &   $1.476 \pm 0.039$  \\ \hline
      $\eta_c N$            &   $0.246 \pm 0.026$   &   $1.703 \pm 0.045$  \\ \hline
    \end{tabular}
  \end{minipage}  
\end{figure}

The effective central potentials are short-ranged, so that they should
receive little finite volume effect. The extrapolation to the infinite
volume can thus be easily done, by fitting the potentials with a
function that goes to zero quickly as r increases. We employ a fit by
two Gaussian functions, $V_{\text{eff}}(r) = \sum_{n=1,2} v_n
\exp(-a_n r^2)$ in all cases.  Then they are used to solve the
effective radial Schr\"odinger equation for the scattering phase
shift.  In Fig.~\ref{fig:phase_shift}, we show the scattering phase
shifts thus calculated as a function of the center-of-mass energy
$E=k^2/(2\mu)$.  No $J/\psi N$ or $\eta_c N$ bound state is observed
below the thresholds. This observation is consistent with the previous
HAL QCD calculations~\cite{Sugiura:2017vks,Kawanai:2010ev} and
another study in the L\"uscher's formalism~\cite{Skerbis:2018lew}.
Interestingly, the NPL collaboration has reported a $\eta_c N$ bound
state with binding energy $19.7\mathrm{MeV}$~\cite{Beane:2014sda} at
the flavor SU(3) point with $m_\pi=807\mathrm{MeV}$, contradicting our
conclusion.
The low-energy S-matrix elements are well parameterized by the
scattering length $a$ and the effective range $r$, defined through the
effective range expansion
$k\cot\delta_0(k)=1/a+rk^2+\mathcal{O}(k^4)$.  Neglecting the
$\mathcal{O}(k^4)$ contributions, we get $a$ and $r$ for the
charmonium-nucleon scattering as in
Table.~\ref{table:scattering_length}.

\section{Conclusions \label{sec:conclusions}}

We have studied the S-wave charmonium-nucleon scattering by using the
time-dependent HAL QCD method. The $J/\psi N$ ($J=1/2$ and $J=3/2$)
and the $\eta_c N$ potentials are all attractive overall, in
qualitative agreement with previous HAL QCD calculations. In this
study, we have found that the $J=1/2$ $J/\psi N$ state obtains
significantly stronger attraction than the $J=3/2$ state. This will be
useful to discuss the applicability of the QCD van der Waals framework
to the charmonium-nucleon systems, since the hyperfine splitting is an
$\mathcal{O}(1/m_c)$ effect and is usually neglected in model
calculations. Quantitative evaluation of the QCD van der Waals
framework is important to see $P_c$ as a $\psi(2S) N$ bound state.
Calculations with lighter pion masses are also necessary for this
purpose.  The coupled-channel analysis for the $\bar{D}^{(*)}
\Sigma_c^{(*)}$ channels is important for a theoretical search of
$P_c$ based on the hadronic molecular picture. This will be left as
our next subject.

\section{Acknowledgements}

This work was supported by JSPS Grant-in-Aid for Scientific
Research~(S), No.~18H0523, Japan Society for the Promotion of Science
KAKENHI Grands No.~JP25400244 and by Ministry of Education, Culture,
Sports, Science and Technology as ``Priority Issue on Post-K
computer'' (Elucidation of the Fundamental Laws and Evolution of the
Universe) and Joint Institute for Computational Fundamental
Science. We thank PACS-CS Collaboration~\cite{PACS-CS} and
ILDG/JLDG~\cite{ILDG_JLDG} for providing the gauge configurations. The
numerical calculations have been performed on Oakforest-PACS at the
University of Tokyo and OCTOPUS in Osaka University.  The lattice QCD
code is partly based on Bridge++~\cite{Bridge}.

\end{document}